\documentclass[aps,pre,showpacs,twocolumn]{revtex4}
\usepackage{amsmath,graphicx,epsfig,amssymb,subfigure,times,dsfont,algorithmic}
\usepackage[usenames]{color}
\usepackage{bbold}

\newcommand{\trace}[1] {\text{Tr}\left[#1\right]}

\newcommand {\Fig}[1] {Fig.~\ref{#1}}
\newcommand {\Eqn}[1] {Eq.~(\ref{#1})}
\newcommand {\Sec}[1] {Sec.~\ref{#1}}

\begin{document}

%\title{Variational Monte Carlo with Unitary Tensor Networks}
\title{Algorithms for the Markov Entropy Decomposition}

\author{Andrew~J.~Ferris}
\author{David~Poulin}
\affiliation{D\'epartement de Physique, Universit\'e de Sherbrooke, Qu\'ebec, J1K 2R1, Canada}

\date{\today}
% blah test
\begin{abstract}
The Markov entropy decomposition (MED) is a recently-proposed, cluster-based simulation method for finite temperature quantum systems with arbitrary geometry. In this paper, we detail numerical algorithms for performing the required steps of the MED, principally solving a minimization problem with a preconditioned Newton's algorithm, as well as how to extract global susceptibilities and thermal responses. We demonstrate the power of the method with the spin-1/2 XXZ model on the 2D square lattice, including the extraction of critical points and details of each phase. Although the method shares some qualitative similarities with exact-diagonalization, we show the MED is both more accurate and significantly more flexible.
\end{abstract}

\pacs{05.10.--a, 02.50.Ng, 03.67.--a, 74.40.Kb}

\maketitle

\section{Introduction}

Although the equations governing quantum many-body systems are simple to write down, finding solutions for the majority of systems remains incredibly difficult. Modern physics finds itself in need of new tools to compute the emergent behavior of large, many-body systems.

There has been a great variety of tools developed to tackle many-body problems, but in general, large 2D and 3D quantum systems remain hard to deal with. Most systems are thought to be non-integrable, so exact analytic solutions are not usually expected. Direct numerical diagonalization can be performed for relatively small systems --- however the emergent behavior of a system in the thermodynamic limit may be difficult to extract, especially in systems with large correlation lengths. Monte Carlo approaches are technically exact (up to sampling error), but suffer from the so-called sign problem for fermionic, frustrated, or dynamical problems. Thus we are limited to search for clever approximations to solve the majority of many-body problems.

Over the past century, hundreds of such approximations have been proposed, and we will mention just a few notable examples applicable to quantum lattice models. Mean-field theory is simple and frequently arrives at the correct qualitative description, but often fails when correlations are important. Density-matrix renormalisation group (DMRG)~\cite{White1992} is efficient and extremely accurate at solving 1D problems, but the computational cost grows exponentially with system size in two- or higher-dimensions~\cite{White1998,Stoudenmire2012}. Related tensor-network techniques designed for 2D systems are still in their infancy~\cite{Verstraete2004,Evenbly2009,Cincio2008}. Series-expansion methods~\cite{OitmaaHamerZheng} can be successful, but may diverge or otherwise converge slowly, obscuring the state in certain regimes. There exist a variety of cluster-based techniques, such as dynamical-mean-field theory~\cite{Georges1996} and density-matrix embedding~\cite{Knizia2012}.

Here we discuss the so-called Markov entropy decomposition (MED), recently proposed by Poulin \& Hastings~\cite{Poulin2011} (and analogous to a slightly earlier classical algorithm~\cite{Globerson2007}). This is a self-consistent cluster method for finite temperature systems that takes advantage of an approximation of the (von Neumann) entropy. In~\cite{Poulin2011}, it was shown that the entropy \emph{per site} can be rigorously upper bounded using only local information --- a local, reduced density matrix on $N$ sites, say.  This approximation becomes exact in the case of a 1D quantum (or classical) Markov chain~\cite{Poulin2011}, and leads to an exponential reduction of cost for exact entropy calculations when the global density matrix is a higher-dimensional Markov network state~\cite{Leifer2008,Brown2012}.

The second approximation used in the MED approach is related to the $N$-representibility problem. Given a set of local but overlapping reduced density matrices $\{\hat{\rho}_i\}$, it is a very challenging problem to determine if there exists a \emph{global} density operator which is positive semi-definite and whose partial trace agrees with each $\hat{\rho}_i$. This problem is QMA-hard (the quantum analogue of NP)~\cite{Liu2006,Liu2007}, and is hopelessly difficult to enforce. Thus, the second approximation employed involves ignoring \emph{global} consistency with a positive operator, while requiring \emph{local} consistency on any overlapping regions between the $\hat{\rho}_i$. At the zero-temperature limit, the MED approach becomes analogous to the variational $n$th-order reduced density matrix approach, where positivity is enforced on all reduced density matrices of size $n$~\cite{Nakata2001,Barthel2012,Baumgratz2012}.

The MED approach is an extremely flexible cluster method, applicable to both translationally invariant systems of any dimension in the thermodynamic limit, as well as finite systems or systems without translational invariance (e.g. disordered lattices, or harmonically trapped atoms in optical lattices). The free energy given by MED is guaranteed to lower bound the true free energy, which in turn lower-bounds the ground state energy --- thus providing a natural complement to variational approaches which upper-bound the ground state energy. The ability to provide a rigorous ground-state energy window is a powerful validation tool, creating a very compelling reason to use this approach.

In this paper we paper we present a pedagogical introduction to MED, including numerical implementation issues and applications to 2D quantum lattice models in the thermodynamic limit. In \Sec{MED}, we give a brief derivation of the Markov entropy decomposition. Section~\ref{Opt} outlines a robust numerical strategy for optimizing the clusters that make up the decomposition. In \Sec{expectation} we show how we can extend these algorithms to extract non-trivial information, such as specific heat and susceptibilities. We present an application of the method to the spin-1/2 XXZ model on a 2D square lattice in \Sec{XXZ}, describing how to characterize the phase diagram and determine critical points, before concluding in \Sec{conclusion}.

\section{Markov entropy decomposition}

\label{MED}

In this section we will briefly review the strategy behind MED, as previously detailed in~\cite{Poulin2011}. We begin in the setting of a quantum lattice model with a Hamiltonian $\hat{H}$ that is `local' according to some graph (e.g. a regular 2D square lattice). For instance, if $\hat{H}$ contains only two-body terms, then we can decompose
\begin{equation}
   \hat{H} = \sum_{<ij>} \hat{h}_{ij},
\end{equation}
where $<\!\!ij\!\!>$ denotes neighboring sites.
The goal is to identify the equilibrium properties of the system at temperature $T$, such as calculating local expectation values (e.g. energy) or determining if the state is in a symmetry-breaking phase.

The global density matrix $\hat{\rho}_G$ describing the equilibrium state at temperature $T$ is the one that minimizes the free energy,
\begin{equation}
   F = E - TS = \trace{\hat{H} \hat{\rho}_G + k_B T \hat{\rho}_G \ln \hat{\rho}_G},
\end{equation}
where $k_B$ is Boltzmann's constant and $\trace{x}$ is the trace of $x$.
The well-known solution to this is given by the Gibbs ensemble,
\begin{equation}
   \hat{\rho}_G = \exp(-\beta \hat{H}) / \trace{\exp(-\beta \hat{H})}, \label{gibbs}
\end{equation}
where $\beta = 1/k_B T$. However, direct evaluation of \Eqn{gibbs} is prohibitively expensive for large systems, primarily because the dimension of $\hat{\rho}_G$ and $\hat{H}$ is simply too large.

In this approach we will store in memory a set of small density matrices $\{\hat{\rho}_i\}$, each corresponding to the reduced state of the system on a set of lattice points (or cluster) $\mathcal{C}_i$. For a consistent description of the state, we must impose that where two clusters $\mathcal{C}_a$ and $\mathcal{C}_b$ partially overlap, the corresponding reduced density matrices on their overlapping region $\mathcal{C}_a \cap \mathcal{C}_b$ must agree. We refer to this condition as \emph{local consistency}.

Given this description of the global state, it is possible to calculate the expectation value of local quantities, such as the individual terms $\langle\hat{h}_{ij}\rangle$ and thus the total energy $E = \langle\hat{H}\rangle$. On the other hand, calculating the system entropy $S = -k_B \trace{\hat{\rho}_G \ln \hat{\rho}_G}$ requires knowledge of the global density matrix.

\begin{figure}
\begin{centering}
\includegraphics[width=0.9\columnwidth]{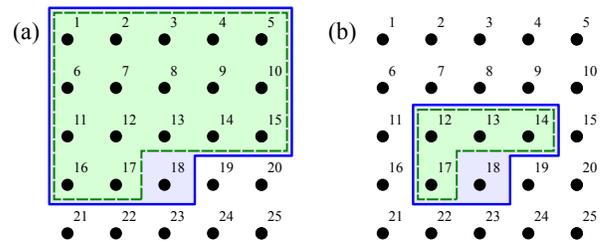}
\end{centering}
\caption{A two-dimension system of 25 spins on a square lattice is depicted, and indexed in a left-to-right then top-to-bottom ``typewriter'' ordering. (a) The quantitity $S(18|1,...17)$ is the difference between the entropies in the region bounded by the blue, solid line, and the region bounded by the green, dashed line. (b) We can approximate this difference using the smaller regions $\mathcal{M}_{19}$ and $\mathcal{M}_{19} \setminus 19$, depicted here, resulting in an upper-bound of the conditional entropy as-per \Eqn{subadd}.  \label{fig:ordering}}
\end{figure}

Nevertheless, we can obtain an approximate value of the system entropy by considering only small segments of the system. We begin by ordering the lattice points according to an integer index $k$. In 1D, the natural ordering is trivial; for 2D systems we use the ``typewriter'' convention of beginning in the top-left, then ordering left-to-right first followed by top-to-bottom (see \Fig{fig:ordering} (a)), and this is simple to generalize to higher dimensions. Note that the choice of ordering is arbitrary --- and each choice will define a unique MED simulation, with possibly different computational cost and accuracy.

Using this ordering, we can decompose the total system entropy into a sum of conditional entropies. Denoting $S(s_1,s_2,\dots)$ as the von Neumann entropy of the reduced state on sites $s_1,s_2,\dots$, we have
\begin{eqnarray}
  S(1,...,N) & = & S(1) + S(1,2) - S(1) + ... \nonumber \\
 & & + \; S(1,...,k) - S(1,...,k-1) + ... \nonumber \\
 & & + \; S(1,...,N) - S(1,...,N-1) \nonumber \\
 & = & \sum_{k=1}^N S(k|1,...,k-1),
\end{eqnarray}
where $S(A|B)$ is the conditional entropy of $A$ given $B$, equal to $S(A,B) - S(B)$. In a typical thermal state, one would expect that the site $k$ would only be strongly correlated with sites that are close to it (according to the interaction graph, say). We consider the set of sites (cluster) $\mathcal{M}_k$ that consists (a) only of sites with index less than or equal to $k$, and (b) are deemed to be in the neighborhood $\mathcal{N}_k$ of site $k$ (including $k$ itself), thus $\mathcal{M}_k = \mathcal{N}_k \cap \{1,...,k\}$. We also define $\mathcal{M}^{\prime}_k = \mathcal{M}_k \setminus k$, the cluster without $k$ itself. In all that follows, the primed symbols will relate specifically to the reduced clusters $\mathcal{M}^{\prime}_k$.

Strong sub-additivity of the von Neumann entropy~\cite{Lieb1973} then gives the inequality
\begin{equation}
    S(k|1,...,k-1) \le  S(k|\mathcal{M}^{\prime}_k). \label{subadd}
\end{equation}
As the cluster $\mathcal{M}_k$ grows to include more of the sites that $k$ is correlated with, the approximation becomes more accurate. In fact, in the special case that the system is a quantum Markov network, we can specify a minimal cluster $\mathcal{M}_k$ such that the entropy decomposition is \emph{exact} (typically, a thermal state is not a quantum Markov network unless the Hamiltonian consists of commuting terms~\cite{Brown2012}). Further, the result can also be seen to be exact~\cite{Poulin2011} when the entropy of the cluster scales simply as $S = \alpha \times \text{volume} + \beta \times \text{boundary} + \gamma$ (where $\alpha$ is a volumetric term coming from thermal fluctuations, the $\beta$ term originates from short-range correlations or entanglement, and the $\gamma$ term is a topological contribution). Such a scaling is the expected fixed point of the renormalization flow for non-critical systems, and thus after sufficient course-graining (or equivalently, growth in the cluster size) the approximation will be quite accurate. 

Equation (\ref{subadd}) implies that the total system entropy is overestimated
\begin{equation}
  S \le  \tilde{S} = \sum_k S(k|\mathcal{M}^{\prime}_k),
\end{equation}
and therefore, we can \emph{under}estimate the free energy of the total system
\begin{equation}
  F \ge  \tilde{F} = E - T\tilde{S} = E - T\sum_k S(k|\mathcal{M}^{\prime}_k).
\end{equation}
using only local, reduced density matrices on the clusters $\mathcal{M}_k$. These relations could be applied in a variety of settings, from estimating the entropy and free energy in an experimental system, to building a numerical technique as we discuss here.

We can attempt to turn the above into a numerical method by minimizing the (approximate) free energy $\tilde{F}$ over the set $\Omega$ of globally consistent density matrices. We say the set of density matrices is globally consistent ($\{\hat{\rho}_k\} \in \Omega$) if and only if there exists a positive, global density operator which agrees with each reduced density matrix after the appropriate partial trace.

Because determining global consistency is extremely difficult, we must make one further approximation in order to proceed. We instead allow any set of operators $\{\hat{\rho}_k\}$ belonging to the enlarged family $\tilde{\Omega} \supset \Omega$ of \emph{locally} consistent operators, which agree on their overlapping sections:
\begin{equation}
   \underset{\mathcal{M}_i \setminus \mathcal{M}_j}{\operatorname{Tr}} \hat{\rho}_i = \underset{\mathcal{M}_j \setminus \mathcal{M}_i}{\operatorname{Tr}} \hat{\rho}_j \label{local}
\end{equation}

The MED algorithm now aims to determine the set of density operators $\{\hat{\rho}_k\} \in \tilde{\Omega}$ that minimizes $\tilde{F}$. Because we have only enlarged our search space ($\tilde{\Omega} \supset \Omega$), this solution also gives a rigorous lower bound to the true free energy,
\begin{equation}
   \min_{\Omega} F \ge \min_{\Omega} \tilde{F} \ge \min_{\tilde{\Omega}} \tilde{F}.
\end{equation}
Further, because the ground state energy $E_0 \ge F$ for all temperatures, we simultaneously have a lower bound for $E_0$.

The full definition of the minimization problem is to minimize the function (where $\hat{\rho}^{\prime}_k = \operatorname{Tr}_k [\hat{\rho}_k]$)
\begin{equation}
    \tilde{F} = \sum_k \operatorname{Tr} \left[ \hat{H} \hat{\rho}_k + k_B T \hat{\rho}_k \log \hat{\rho}_k \right] - \operatorname{Tr} \left[k_B T \hat{\rho}^{\prime}_k \log \hat{\rho}^{\prime}_k \right] \label{F},
\end{equation}
over the set of density matrices that are Hermitian, positive semi-definite ($\hat{\rho}_k \succeq 0$) with unit trace, and are locally consistent (i.e. obey \Eqn{local}). At $T = 0$ this reduces to a much easier semi-definite program (for examples, see Refs.~\cite{Nakata2001,Barthel2012,Baumgratz2012}). Note that $\tilde{F}$ may not be monotonic with $T$, and better bounds on $E_0$ may be found with the MED at finite temperature, as can be clearly seen in \Fig{fig:compare_syms}

Note that this approach can easily be applied to infinite, translationally invariant systems. If the system is translationally invariant, the reduced density operator will appear the same for all $k$ (note that even in a symmetry-breaking phase, the thermal state is a mixture of all sectors). Thus each term in \Eqn{F} is equal, and it suffices to minimize over just one, locally consistent density matrix. On the other hand, for non-translationally invariant systems containing $N$ sites (or a unit-cell of $N$ sites), a complete set of $N$ clusters should be used.  Care should be taken at the boundaries of a finite, translationally invariant system (e.g. the first cluster always contains just one site, even for periodic systems).

Before we continue to the specifics of numerical optimization in \Sec{Opt}, we wish to point out one of the general properties of MED: the total minimization problem is convex (\Eqn{F} is convex, the set of unit-trace, positive density operators is convex, while the local consistency conditions induce an linear subspace that the search is constrained within). This implies that there is just one unique minima to the problem, and as the objective function [\Eqn{F}] is smooth, robust and reliable numerical techniques allow us to determine the minima to great precision. This is particularly important when claiming a lower-bound to the free energy --- a claim which relies on an accurate minimization.

\section{Optimization}

\label{Opt}

In this section we will discuss numerical strategies for minimizing \Eqn{F}. To solve this constrained optimization problem, we use a combination of interior-point potentials and subspace projection, as discussed in \Sec{Opt:Con}. We will use the first- and second-derivatives to minimize \Eqn{F}; numerically efficient strategies for calculating these are presented in \Sec{Opt:Der}. In \Sec{Opt:Alg} we present reliable convex optimization algorithms based on nonlinear conjugate-gradient or Newton's method.

\subsection{Constraints}

\label{Opt:Con}

The minimization problem is constrained in two ways. Firstly, each $\hat{\rho}_i$ must be (semi-) positive-definite. Secondly, the density matrices must be locally consistent, and have unit trace. The latter constraints limit the search space to an affine subspace, discussed further below.  To ensure we satisfy the former, we include an interior-point (IP) potential~\cite{BoydVandenberghe}, proportional to the control variable $\mu$, into the cost function:
\begin{eqnarray}
    \tilde{F}  = \sum_k & \mathrm{Tr} \left( \hat{H} \hat{\rho}_k + k_B T \hat{\rho}_k \log \hat{\rho}_i + \mu \log \hat{\rho}_k  \right) \nonumber & \\
     & - \; \mathrm{Tr} \left(k_B T \hat{\rho}^{\prime}_k \log \hat{\rho}^{\prime}_k \right). & \label{F_IP}
\end{eqnarray}
The additional term creates a logarithmic potential barrier surrounding the surface of singular density matrices. The logarithmic form is by far the most commonly used IP potential, being sufficiently strong to ensure stability while decaying extremely quickly away from this surface. In this case, it also preserves the convexity of the cost function. By carefully choosing $\mu$, and reducing it appropriately, we can converge to the (unique) minimum within the space of positive density matrices, while removing instability issues in nearly-singular regions.

The remaining constraints impose that local consistency (\Eqn{local}) and the correct normalization,
\begin{equation}
  \mathrm{Tr} \, \hat{\rho}_i = 1. \label{trace}
\end{equation}
Equation (\ref{local}) can be written more conveniently in the form
\begin{equation}
    \underset{\mathcal{M}_i \setminus \mathcal{M}_j}{\operatorname{Tr}} \hat{\rho}_i - \underset{\mathcal{M}_j \setminus \mathcal{M}_i}{\operatorname{Tr}} \hat{\rho}_j = 0\label{consistency}
\end{equation}
Equations~(\ref{trace},\ref{consistency}) are first-degree polynomials in the search-space variables (i.e. the elements of $\hat{\rho}_i$), and define an affine subspace. It is well-known that imposing affine constraints onto a convex optimization problem preserves convexity~\cite{BoydVandenberghe}.

The space of differences between two normalized, locally-consistent states is \emph{linear}. Thus, we can satisfy Eqs.~(\ref{trace},\ref{consistency}) exactly by beginning with a locally-consistent state (e.g. the fully mixed state), and projecting all updates $\Delta \hat{\rho}_i$ to the variables onto this linear subspace. The remainder of this subsection is devoted to how to construct the appropriate orthogonal projector correctly and efficiently.

Formally, we wish to construct a projector $\mathcal{P}$ such that $\mathcal{P}\{\hat{x}_1,...,\hat{x}_N\}$ returns the closest (according to the $L_2$-norm) set of numbers $\{\hat{y}_1,...,\hat{y}_N\}$ satisfying
\begin{equation}
  \mathrm{Tr} \, \hat{y}_i = 0, \label{trace2}
\end{equation}
\begin{equation}
    \underset{\mathcal{M}_i \setminus \mathcal{M}_j}{\operatorname{Tr}} \hat{y}_i - \underset{\mathcal{M}_j \setminus \mathcal{M}_i}{\operatorname{Tr}} \hat{y}_j = 0. \label{consistency2}
\end{equation}
One method of performing this would be to simply construct the (sparse) super-operator $\mathcal{P}$ in memory and perform the projection as a matrix-vector multiplication. However, the dimension of the super-operator is too large for this method to be efficient. Here we discuss how to perform this projection in a step-wise fashion.

Firstly, the traceless constraint \Eqn{trace2} is orthogonal to the remaining constraints and thus can be enforced independently (simply by calculating the complete trace, and subtracting the appropriate identity operator). As different terms in \Eqn{consistency2} are generally not orthogonal constraints, there are some subtleties in the projection algorithm.

\begin{figure}
\begin{centering}
\includegraphics[width=0.65\columnwidth]{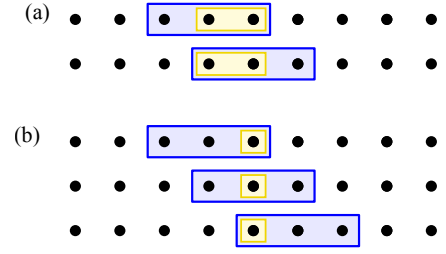}
\end{centering}
\caption{A cluster of three sites on a translationally invariant chain must obey certain local consistency conditions. (a) The two-site condition $\hat{\rho}_{12} = \hat{\rho}_{23}$ arises from translating the cluster by just one site. (b) The one-site conditions $\hat{\rho}_1 = \hat{\rho}_2 = \hat{\rho}_3$ can be seen from two separate translations of the cluster. Note that although (b) is a direct consequence of (a), both must be taken into account when constructing the action of the projector $\mathcal{P}$. \label{fig:consistency}}
\end{figure}

To demonstrate this subtlety, we can try to enforce a single constraint of the form \Eqn{consistency2}, for instance the two-body term in \Fig{fig:consistency}~(a). Here we are working in the framework of translationally invariant systems, but the same principles hold for generic systems. Given an operator $\hat{\rho}_{123}$ (on sites labeled 1,2,3), we wish to impose $\hat{\rho}_{12} = \hat{\rho}_{23}$ (taking the appropriate partial traces). As a first-guess, we can try enforcing that these two operators become their average, $(\hat{\rho}_{12} + \hat{\rho}_{23})/2$:
\begin{equation}
  \hat{\rho}^{\prime}_{123} = \hat{\rho}_{123} + \frac{-\hat{\rho}_{12} + \hat{\rho}_{23}}{2} \otimes \frac{\hat{\mathbb{1}}_3}{d} + \frac{\hat{\mathbb{1}}_1}{d} \otimes \frac{\hat{\rho}_{12} - \hat{\rho}_{23}}{2}.
\end{equation}
However, the problem with this is that $\hat{\rho}^{\prime}_{12} \ne \hat{\rho}^{\prime}_{23}$ \emph{unless it is already true that} $\hat{\rho}_1 = \hat{\rho}_2 = \hat{\rho}_3$ (see \Fig{consistency}~(b)). So, even though this one-site constraint is actually implied by the two-site constraint, we must enforce the one-site constraint manually first. Observe that the algorithm:
\begin{eqnarray}
   \hat{\rho}^{\prime}_{123} & = & \hat{\rho}_{123} + \frac{-2\hat{\rho}_{1} + \hat{\rho}_{2} + \hat{\rho}_{3}}{3} \otimes \frac{\hat{\mathbb{1}}_{23}}{d^2} \nonumber \\
   & & + \; \frac{\hat{\mathbb{1}}_1}{d} \otimes \frac{\hat{\rho}_{1} - 2\hat{\rho}_{2} + \hat{\rho}_{3}}{3} \otimes \frac{\hat{\mathbb{1}}_3}{d} \nonumber \\
   & & + \; \frac{\hat{\mathbb{1}}_{12}}{d^2}  \otimes \frac{\hat{\rho}_{1} + \hat{\rho}_{2} - 2\hat{\rho}_{3}}{3}, \nonumber \\
   \hat{\rho}^{\prime\prime}_{123} & = & \hat{\rho}^{\prime}_{123} + \frac{-\hat{\rho}^{\prime}_{12} + \hat{\rho}^{\prime}_{23}}{2} \otimes \frac{\hat{\mathbb{1}}_3}{d} \nonumber \\
   & & + \; \frac{\hat{\mathbb{1}}_1}{d} \otimes \frac{\hat{\rho}^{\prime}_{12} - \hat{\rho}^{\prime}_{23}}{2} , \label{three_site_constraints}
\end{eqnarray}
performs an orthogonal projection that enforces $\hat{\rho}^{\prime\prime}_{12} = \hat{\rho}^{\prime\prime}_{23}$.

The projection algorithm proceeds as follows. First we identify any one-site constraints, which are enforced. Then we enforce the two-site constraints, followed by three-site constraints, and so forth. The problem now falls to identifying all the $n$-site constraints of the problem, and carefully discarding any that are unnecessary.

\begin{figure}
\begin{centering}
\includegraphics[width=\columnwidth]{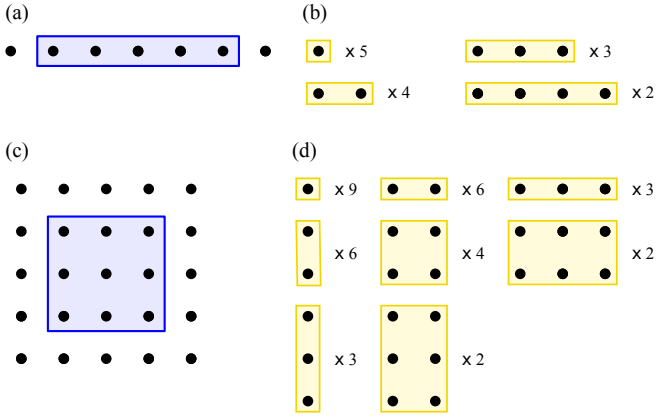}
\end{centering}
\caption{Example clusters and their relevant sub-cluster consistency conditions. (a) A cluster of 5 sites on the translationally-invariant, infinite chain. (b) These four sub-clusters are required to enforce local consistency, construct the projector $\mathcal{P}$. The number of unique translations where the sub-cluster fits entirely in the original cluster (a) is marked beside each. (c) A $3\times3$ cluster on the square lattice. (d) In this case, all rectangular sub-clusters are required to construct the projector $\mathcal{P}$. \label{fig:consistency2}}
\end{figure}

Take for instance a contiguous cluster of $N$-sites on a 1D, translationally-invariant chain, as in \Fig{fig:consistency2}~(a). Then the minimal set of sub-constraints to project into the translational invariant subspace consists of all the contiguous sub-clusters of size $1,...,N-1$ (see \Fig{fig:consistency2}~(b)). Any constraints on non-contiguous sub-clusters are implied by (but not required by) the smallest enveloping sub-cluster. The projection $\mathcal{P}\{\hat{\rho}\}$ is performed by a series of $N-1$ steps, similar to \Eqn{three_site_constraints}.

\begin{figure}
\begin{centering}
\includegraphics[width=0.8\columnwidth]{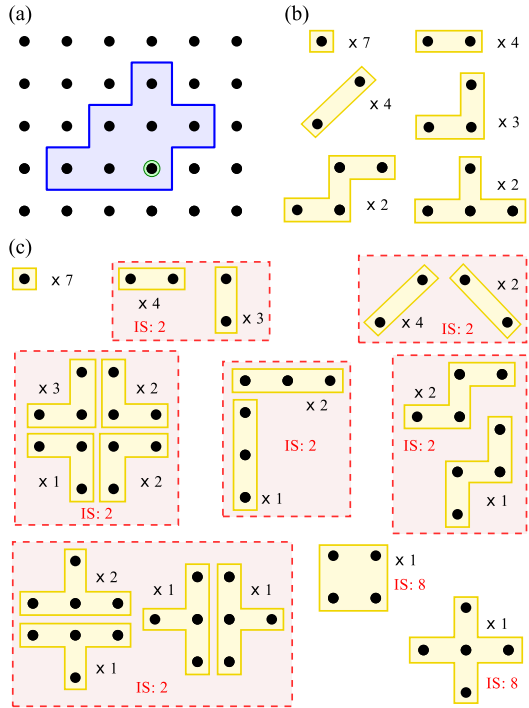}
\end{centering}
\caption{A more complicated cluster and its relevant sub-cluster consistency conditions. (a) An cluster of 7 sites on the square lattice. This cluster $\mathcal{M}_k$ may be used to estimate the entropy of the highlighted site $k$ conditioned on sites with index $< k$, i.e. $S(k | \{ < k \} )$. (b) Six sub-clusters are required to enforce translational invariance. (c) Sub-clusters for translational, rotational and reflection symmetries. Sub-clusters related to each other by rotation and/or reflection should be dealt with simultaneously. Some sub-clusters have internal rotational or reflect symmetries --- the total number of internal symmetries is labeled by IS. \label{fig:consistency3}}
\end{figure}

The solution in higher-dimensions may be more complicated. For a regular square lattice, translational-invariance on a rectangular cluster requires consecutively imposing the appropriate constraint on all the rectangular sub-clusters, in order of size (see \Fig{fig:consistency2}~(c,d)). In general, each element of the translation group may generate a sub-cluster constraint; by taking the original cluster and translated copy, the overlapping region corresponds to a constraint. A relevant example is depicted in \Fig{fig:consistency3}~(a,b). The projection $\mathcal{P}$ requires six distinct steps involving the sub-clusters in \Fig{fig:consistency3}~(b)

\begin{figure}
\begin{centering}
\includegraphics[width=0.75\columnwidth]{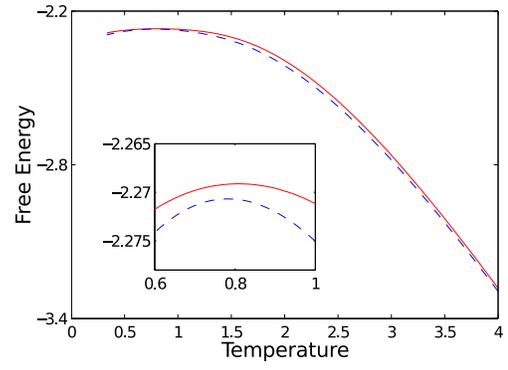}
\end{centering}
\caption{MED predictions of the free energy for the antiferromagnetic, spin-1/2 XY model (Hamiltonian \Eqn{H} with $\Delta = 0$, $J = 1$) on the square lattice, using the 7-site cluster depicted in \Fig{fig:consistency3}. The solid, red line corresponds to results when enforcing all point symmetries of the square lattice, while the dashed, blue line comes from simulations enforcing only translational invariance. The larger symmetry group results in a strictly greater free energy, and a higher bound on the ground state energy ($E_0 \ge -2.26912$ or $-2.27068$, respectively). %TODO: Compare these results to variational and MC studies 
\label{fig:compare_syms}}
\end{figure}

For Hamiltonians with a larger symmetry group (such as rotational or reflection symmetry), it becomes possible to investigate all sub-clusters generated by the total symmetry group (and the closure of this originating from overlaps between sub-clusters --- see \Fig{fig:consistency3}~(c)). This will lead to a larger set of constraints to impose, but improves the accuracy of the MED approximation because the set of allowed density matrices $\Omega^{\prime}$ is reduced (see the comparison in \Fig{fig:compare_syms}). Occasionally, this procedure generates a sub-cluster constraint that is a direct descendent of another, and is therefore unnecessary to apply (detecting these cases is simple by direct inspection).

The construction of the projector $\mathcal{P}$ is just one major element of our optimization procedure. In the next section we use $\mathcal{P}$ to calculate derivatives of $F^{\prime}$ within the constrained subspace.

\subsection{Calculating the Derivatives}

\label{Opt:Der}

The optimization will proceed by treating the elements of the density matrix as a list of variables, which we vary to find the (unique) free energy minimum. The first- and second-derivatives of the free energy with respect to these variables can be very helpful in determining how to vary these variables to approach the minimum. The results of matrix calculus~\cite{Higham,matrixcookbook} will allow us to find the derivatives of $\tilde{F}$ w.r.t. $\hat{\rho}_i$.

The first derivative is relatively straightforward to compute.
\begin{equation}
    \frac{d\tilde{F}}{d\hat{\rho}_i} = \hat{H} + k_B T \left ( \log \hat{\rho}_i - \frac{\hat{\mathbb{1}}}{d} \otimes \log \hat{\rho}^{\prime}_i \right) + \mu \hat{\rho}_i^{-1} \label{dF}
\end{equation}
Recall that the primed variables $\hat{\rho}^{\prime}_k$ correspond to the reduced density matrix on the cluster \emph{excluding} $k$ itself, i.e. $\hat{\rho}^{\prime}_k = \text{Tr}_k \, \hat{\rho}_k$. 

However, this does not take into account that the variables are \emph{constrained} to the locally-consistent subspace. The projected derivative
\begin{equation}
    \nabla \tilde{F} = \mathcal{P} \left\{ \frac{\partial \tilde{F}}{\partial \hat{\rho}_1},\dots, \frac{\partial \tilde{F}}{\partial \hat{\rho}_N} \right\} \label{proj_dF}
\end{equation}
lives entirely in the correct subspace, returning the same inner-product on locally-consistent displacements. The projected derivative is efficient to compute, typically by directly diagonalizing $\hat{\rho}_k$ and $\hat{\rho}^{\prime}_k$ to calculate the matrix logarithm. The total cost scales as $\mathcal{O}(N d^{3n})$, where $N$ is the number of distinct clusters (just one, for translationally invariant systems in the thermodynamic limit), $d$ is the Hilbert space dimension of each lattice site, and $n$ is the number of sites per cluster.

Using the projected derivative, we can minimize $\tilde{F}$ using a projected-subspace, derivative-based method, such as the steepest-descent, nonlinear conjugate-gradient or L-BFGS algorithms. Conjugate gradient was used in Ref.~\cite{Poulin2011} to minimize the free energy, however as it was observed there, accurate minimization is challenging, while convergence is quite slow. We explain that this behaviour is due to stiffness, below.

Using the results of matrix calculus (see the Appendix) we obtain the second-derivative, or Hessian super-operator, which has the form
\begin{equation}
    \frac{\partial^2 \tilde{F}}{\partial\hat{\rho}_i \partial\hat{\rho}_j} = 0, \label{hessian0}
\end{equation}
when $i \ne j$, and
\begin{eqnarray}
    \frac{\partial^2 \tilde{F}}{\partial\hat{\rho}_i\vphantom{.} ^2} & = & \hat{U}^{\dag}_i \otimes \hat{U}_i^{T} \cdot \mathrm{diag} \; \hat{M}_i \cdot \hat{U}_i \otimes \hat{U_i}^{\ast}  \\
    & - & \mathfrak{ex} \cdot \hat{U}^{\prime\dag}_i \otimes \hat{U}_i^{\prime T} \cdot \mathrm{diag} \; \hat{M}^{\prime}_i \cdot \hat{U}^{\prime}_i \otimes \hat{U_i}^{\prime\ast} \cdot \mathfrak{pt}, \nonumber \label{hessian1}
\end{eqnarray}
where $\hat{\rho}_i = \hat{U}_i^{\dag} \hat{D}_i \hat{U}_i$ and $\hat{\rho}^{\prime}_i = \hat{U}^{\prime \dag}_i \hat{D}^{\prime}_i \hat{U}^{\prime}_i$ are the eigenvalue decompositions of $\hat{\rho}_i$ and $\hat{\rho}^{\prime}_i$, while $\mathrm{diag}(\hat{x})$ is the diagonal super-operator that multiplies element-wise by the entries of $\hat{x}$. The diagonals of matrix $\hat{M}_i$ is defined as
\begin{equation}
    (\hat{M}_i)_{jj} = k_B T (\hat{D}_i)_{jj}^{-1} - \mu (\hat{D}_i)_{jj}^{-2} , \label{hessian1a}
\end{equation}
and the off-diagonals
\begin{equation}
    (\hat{M}_i)_{jk} =  \frac{k_B T (\ln(\hat{D}_i)_{jj} - \ln (\hat{D}_i)_{kk}) - \mu((\hat{D}_i)_{jj}^{-1} - (\hat{D}_i)_{kk}^{-1})}{(\hat{D}_i)_{jj} - (\hat{D}_i)_{kk}}. \label{hessian1b}
\end{equation}
$\hat{M}_i^{\prime}$ is defined similarly based on the values of $\hat{D}_i^{\prime}$:
\begin{equation}
   (\hat{M}^{\prime}_i)_{jj} = k_B T (\hat{D}^{\prime}_i)_{jj}^{-1} , \label{hessian1c}
\end{equation}
\begin{equation}
   (\hat{M}^{\prime}_i)_{jk}  = k_B T \frac{\ln(\hat{D}^{\prime}_i)_{jj} - \ln (\hat{D}^{\prime}_i)_{kk}}{(\hat{D}^{\prime}_i)_{jj} - (\hat{D}^{\prime}_i)_{kk}}. \label{hessian1d}
\end{equation}
Finally, the partial-trace super-operator $\mathfrak{pt}$ performs the partial trace over site $k$, for example, $\hat{\rho}^{\prime}_k = \text{Tr}_k \, \hat{\rho}_k = \mathfrak{pt} \, \{ \hat{\rho}_k \}$. The quasi-inverse of this operation, the expansion super-operator $\mathfrak{ex}$, adds the completely mixed site at position $k$, so $\mathfrak{ex} \, \{\hat{\rho}^{\prime}_k\} = \hat{\rho}^{\prime}_k \otimes \hat{\mathbb{1}} / d$.

The Hessian on the locally-consistent subspace can be found by simply applying the projection super-operator before and after the second-derivative.
\begin{equation}
    \nabla^2 \tilde{F} = \mathcal{P} \cdot \mathrm{diag} \left\{\frac{\partial^2 \tilde{F}}{\partial \hat{\rho}_1^2}, \dots, \frac{\partial^2 \tilde{F}}{\partial \hat{\rho}_N^2} \right\}\cdot \mathcal{P} \label{hessian2}
\end{equation}
In this equation, diag is the expansion into a matrix over the set of clusters $i,j$ (representing the diagonal structure given by \Eqn{hessian0}).

Overall, the \emph{action} of the Hessian is efficient to compute, requiring time scaling as $\mathcal{O} (N d^{3n})$, similar to the derivative. Thus, it is feasible to use Hessian-based optimization strategies, particularly given that the diagonalization of $\hat{\rho}_k$ and $\hat{\rho}^{\prime}_k$ may be the most expensive operations. In the next section we discuss a preconditioned Newton's algorithm which we have implemented here.

Finally, we wish to discuss the issue of stiffness in this minimization problem. By inspection of the unprojected Hessian, Eqs.~(\ref{hessian1}--\ref{hessian1d}), we can determine a simple scaling of the stiffness as a function of $\hat{\rho}_k$. To begin with, the first term in \Eqn{hessian1} dominates (when $\mu = 0$); it is positive-definite (as entropy is convex), while the second term is negative-definite (for the same reason), and the sum is known to be positive semi-definite (because $\tilde{F}$ is convex). Thus, the subtracted term will typically \emph{increase} the stiffness of the Hessian (this can be proven in the case that $\hat{\rho}_k$ and $\hat{\rho}^{\prime}_k$ can be co-diagonalized). The first term is simply the matrix derivative of $\ln \hat{\rho}_k$, which very roughly speaking scales like $\hat{\rho}_k^{-1}$. Therefore, we expect the stiffness of the problem to be related to the ratio of eigenvalues in $\hat{\rho}_k$, which for thermal states increases exponentially with inverse temperature $\beta$. It is in-fact the very small eigenvalues of $\hat{\rho}_k$ that lead to the minimization difficulties seen in Ref.~\cite{Poulin2011}. Fortunately, the interior-point potential proportional to $\mu$ ameliorates this problem signficantly by artificially increasing the small eigenvalues of $\hat{\rho}_k$. At low temperatures we expect that these small eigenvalues, and thus the stiffness, scale like $\mathcal{O} (1/\mu T)$.

\subsection{Optimization algorithm}

\label{Opt:Alg}

We have made the following three major observations: (a) the minimization problem is convex, (b) the action of the Hessian matrix is relatively efficient to compute, and (c) the minimization problem can be very stiff. Given these ingredients, Newton's method is a natural and simple choice for accurately determining the minimum.

The updates in Newton's method are straightforward to define:
\begin{equation}
    \hat{\rho}_i := \hat{\rho}_i - h \left(\nabla^2 \tilde{F}\right) ^ {-1} \left\{ \nabla \tilde{F} \right\}, \label{Newton}
\end{equation}
where $h$ will usually equal 1, except when the step-size is detected to be too great, in which case a simple backtracking line-search is utilized~\cite{NumericalRecipes}. As it is too expensive to construct and invert the full Hessian matrix, a (linear) conjugate-gradient algorithm can be used to determine $\left(\nabla^2 \tilde{F}\right) ^ {-1} \left\{ \nabla \tilde{F} \right\}$. However, as the Hessian matrix is quite stiff, preconditioning can reduce the number of conjugate gradient steps dramatically.

A preconditioner can be constructed by finding a matrix which is an approximation to the inverse of the Hessian. In general, the projection $\mathcal{P}$, and each of the two terms in \Eqn{hessian1}, do not commute. As a simple approximation, we invert the positive term in \Eqn{hessian1} (i.e. ) and apply the projection as-if it did commute:
\begin{equation}
    \mathcal{T} = \mathcal{P} \cdot \left\{ \mathrm{diag} \; T_i \right\}\cdot \mathcal{P} \label{precon1}
\end{equation}
Here, $T_i$ is simply the inverse of $M_i$.
\begin{equation}
    T_i=  \hat{U}^{\dag}_i \otimes \hat{U}_i^{T} \cdot \mathrm{diag} \; \hat{t}_i \cdot \hat{U}_i \otimes \hat{U_i}^{\ast}  \label{precon2}
\end{equation}
\begin{equation}
    (\hat{t}_i)_{jj} = \frac{1}{k_B T (\hat{D}_i)_{jj}^{-1} - \mu (\hat{D}_i)_{jj}^{-2}} , \label{precon3a}
\end{equation}
\begin{equation}
    (\hat{t}_i)_{jk} =  \frac{(\hat{D}_i)_{jj} - (\hat{D}_i)_{kk}}{k_B T (\ln(\hat{D}_i)_{jj} - \ln (\hat{D}_i)_{kk}) - \mu((\hat{D}_i)_{jj}^{-1} - (\hat{D}_i)_{kk}^{-1})}. \label{precon3b}
\end{equation}

Note that this preconditioner is also suitable for a preconditioned \emph{nonlinear} conjugate gradient or L-BFGS search. We found that this preconditioner significantly accelerates the inner-CG solver in our Newton's method implementation. However, the preconditioned Hessian $\mathcal{T} \cdot \nabla^2 \tilde{F}$ remains quite stiff at low temperatures --- numerical investigation leads us to believe that the approximate treatment of the projector $\mathcal{P}$ has a larger impact than neglecting the second term in \Eqn{hessian1}.

As the optimization progresses, we can estimate the error in our solution using the Hessian matrix, or its approximate inverse $\mathcal{T}$. Thus we can use $\nabla \tilde{F} \cdot (\nabla^2 \tilde{F})^{-1} \{\nabla \tilde{F}\}$ for an determination of the proximity to the true minimum, or $\nabla \tilde{F} \cdot \mathcal{T} \{\nabla \tilde{F}\}$ for a rough estimation. If sufficient CG steps are taken, we observe super-linear convergence.

Because our stopping criteria depend on the derivatives, this minimization can proceed well beyond the numerical accuracy of $F$ itself, leading to a very accurate determination of the density matrices $\hat{\rho}_i$, as well as very rigorous bounds on $F$ and $E_0$ (extrapolation as $\mu \rightarrow 0$ may be necessary before claiming lower bounds).

Finally, we wish to point out one possible way to improve our optimization strategy. As an alternative approach to the interior-point potential to maintain positivity, one could use a positive representation, such as the square-root of the density matrices,
\begin{equation}
    \hat{\rho}_i = \hat{A}^{\dagger}_i \hat{A}_i.
\end{equation}
This decomposition is not unique, but we could constrain $\hat{A}_i$ to be Hermitian. Optimization could then proceed over the variables $\hat{A}_i$. A similar analysis of the Hessian matrix to above leads us to believe that the optimization problem becomes less stiff. On the other hand, the local consistency constraints are now quadratic polynomials of the search-space variables, and so a Langrange-multiplier approach may be necessary. Also note that small eigenvalues of $\hat{\rho}_i$ could be removed entirely by using rectangular matrices $\hat{A}_i$.

\section{Expectation values and susceptibility}

\label{expectation}

Before proceeding to numerical simulations in \Sec{XXZ}, we discuss how to determine physical quantities of interest from the MED. Although we have explicit access to local information, stored in the reduced density matrices $\hat{\rho}_i$, more sophisticated techniques yield global properties such as the magnetic susceptibility or specific heat.

The presence of long-range order is not simple to detect in the MED formalism. To begin with, we do not have an algorithm for determining long-distance correlations (except in 1D, where belief propagation~\cite{Hastings2007} may be appropriate). In a symmetry-broken phase, the free energy is minimized by the equal mixture of the relevant symmetry sectors. Further, this mixed state is the unique minimum of the MED optimization problem, because the cost functional $F$ (\Eqn{F}) is both convex and fully analytic (smooth). Therefore, we do not expect to observe either long-range order or explicit symmetry breaking. In fact, because $F$ is always smooth, we do not expect to observe phase transitions at all!

This property is similar to exact-diagonalization of small systems; in-fact MED and exact diagonalization share many similarities, including scaling of numerical cost with $n$ and strong finite-size effects. Nonetheless, we show here how to extract signatures of many-body cooperation and phase transitions in the thermodynamic limit, such as diverging magnetic susceptibility or specific heat. In \Sec{XXZ}, we demonstrate that the many-body effects are much stronger in the MED than exact diagonalization, even when using significantly smaller clusters.

The specific heat $dE/dT$ could be calculated by numerical derivative of the energy at different temperatures; however it is also possible to determine directly from the equilibrium state at temperature $T$. The solution to the MED minimization always satisfies
\begin{eqnarray}
   \nabla \tilde{F} = 0 & = & \mathcal{P} \biggl\{ \hat{H}_1 + k_B T \left ( \log \hat{\rho}_1 - \frac{\hat{\mathbb{1}}}{d} \otimes \log \hat{\rho}^{\prime}_1 \right) \\ \nonumber
        & & + \; \mu \hat{\rho}_1^{-1}, \dots \biggr\}.
\end{eqnarray}
An infinitesimal change in temperature, $T \rightarrow T + dT$, will result in a small change in the solution $\hat{\rho}_i \rightarrow \hat{\rho}_i + d\hat{\rho}_i$. Inserting this into the above and rearranging the differentials to find an equation for $d\hat{\rho}_i/dT$ yields
\begin{equation}
    \nabla^2 \tilde{F} \left\{ \frac{d\hat{\rho}_1}{dT}, \dots \right\} = \frac{1}{T} \mathcal{P} \, \left\{ \hat{H}_1,\dots \right\} .
\end{equation}
Although we have not determined an analytic solution, in \Sec{Opt:Alg} we introduced an efficient preconditioned-CG algorithm for inverting the action of $\nabla^2 \tilde{F}$. It becomes straightforward to measure the thermal response at the end of each minimization. We can then simply calculate how any local quantity responds to the change in temperature, e.g. the specific heat
\begin{equation}
   c_V = \frac{dE}{dT} = \sum_i \trace{\hat{H}_i \frac{d\hat{\rho}_i}{dT}}.
\end{equation}

This procedure can be generalized to the response with respect to perturbations to the Hamiltonian. Take, for instance, the addition of a small, global magnetic field, such that $\hat{H}_i \rightarrow \hat{H}_i + dB \hat{\sigma}^z_i$. Similar manipulations to the above yield
\begin{equation}
    \nabla^2 \tilde{F} \left\{ \frac{d\hat{\rho}_1}{dB}, \dots \right\} = - \mathcal{P} \, \left\{ \hat{\sigma}^z_1,\dots \right\},
\end{equation}
while the magnetic susceptibility is simply
\begin{equation}
   \chi = -\frac{d\langle\hat{\sigma}^z\rangle}{dB} = -\sum_i \trace{\hat{\sigma}^z_i \frac{d\hat{\rho}_i}{dB}}.
\end{equation}
We observe in the next section that the magnetic susceptibility calculated by MED can be many orders of magnitude greater than that of exact diagonalization in ferromagnetic symmetry-broken phases, while using less sites in the cluster.

This technique is quite powerful and flexible. Take, for example, a simple antiferromagnetic phase. The MED will return a mixture of symmetry sectors, yielding no breaking of translational invariance, and the minimization can be performed using just one cell. The staggered (or sublattice) magnetization must be, by definition, zero in this translationally invariant ansatz. On the other hand, the staggered magnetic susceptibility is defined by the response to adding a perturbative, staggered magnetic field to the Hamiltonian, which explicitly breaks the original translational-invariance. For efficiency, we can \emph{transfer} the zero-field solution to the checkerboard lattice, and then add a perturbative, staggered magnetic field to the Hamiltonian, repeating the above procedure to calculate the staggered magnetic susceptibility. Thus, the expensive step of minimization can always be performed using the full translational invariance of the Hamiltonian, only breaking this for the relatively cheaper operation of determining the response to small fields. This can be extended to cells much larger than the MED clusters themselves, such as phases with chiral-order or similar.

\section{XXZ Model}

\label{XXZ}

In this section we benchmark the MED approach with the XXZ model on the 2D square lattice, and compare the results to those obtained with exact diagonalization of small systems with periodic boundary conditions. Unsurprisingly, we find that both the MED and exact-diagonalization approaches display pronounced finite-size effects. We observe that the MED displays similar accuracy and sharpness of features using approximately half the number of sites, compared to exact diagonalization, while at the same time being much more flexible.

The Hamiltonian under consideration is
\begin{equation}
 \hat{H} = J \sum_{<ij>} \Delta \hat{\sigma}^z_i \hat{\sigma}^z_j + \left( \hat{\sigma}^x_i \hat{\sigma}^x_j + \hat{\sigma}^y_i \hat{\sigma}^y_j \right), \label{H}
\end{equation}
where $<\!\!ij\!\!>$ denotes pairs of neighboring sites on a 2D square lattice. This model has a rich phase diagram containing several %second-order 
classical (i.e finite temperature) and quantum (i.e. zero temperature) phase transitions.

For the purpose of displaying the entire phase diagram, we rewrite the above in the form
\begin{equation}
 \hat{H} = \sum_{<ij>} \cos(\pi\alpha) \hat{\sigma}^z_i \hat{\sigma}^z_j + \sin(\pi\alpha) \left( \hat{\sigma}^x_i \hat{\sigma}^x_j + \hat{\sigma}^y_i \hat{\sigma}^y_j \right), \label{H2}
\end{equation}
where the anisotropy is given in terms of $\alpha$ by $\cot(\pi\alpha) = \Delta$, while the coupling strength $J$ is fixed by $\sin(\pi\alpha) = J$.

Although no general analytical solution exists, certain values of $\alpha$ correspond to well-understood systems. At $\alpha = 0$ we recover the classical antiferromagnetic (AFM) Ising model, while $\alpha = 1$ corresponds to the ferromagnetic (FM) Ising model. The AFM and FM quantum XX model ($\Delta = 0$) is given at $\alpha = 0.5$ and $1.5$, respectively. The AFM and FM Heisenberg model ($\Delta = 1$) is given at points $\alpha = 0.25$ and $1.25$, while a local unitary transformation maps $\alpha$ to $2 - \alpha$. The ``Ising-like" FM phase with $0.75 \le \alpha \le 1.25$ have known, product-state ground states, with all spins pointing in the same direction.

The Hamiltonian has a $U(1)$ symmetry corresponding to global rotations of the spin-vector about the $z$-axis, or equivalently the conservation of the $z$-projection of the total spin. This symmetry can be exploited numerically to decrease time and memory requirements, by storing the density-matrices $\hat{\rho}_i$ in block-diagonal form.

\begin{figure}[t]
    \begin{centering}
    \includegraphics[width=8cm]{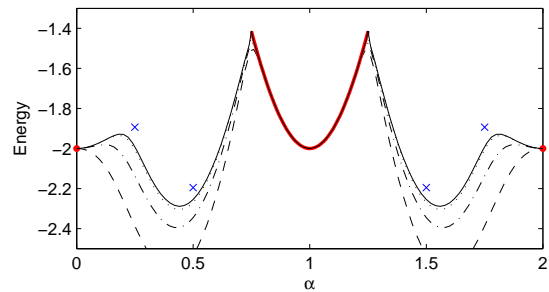}
    \caption{(Color online) Ground state energies of the XXZ model, as a function of $\alpha$. The black lines corresponds to the MED results with clusters of different sizes. From top-to-bottom, we use an 8-site cluster (solid line), 7-site cluster (dotted line), 5-site cluster (dash-dot line) and 3-site cluster (dashed line). The solid line forms a rigorous lower-bound to the ground state energy. For comparison, the thick, red line and red points correspond to the exact product-state ground states, while the blue crosses represent Monte Carlo results for the Heisenberg~\cite{Sandvik1999} and XY~\cite{Sandvik1997} AFMs respectively. \label{fig:ground}}
    \end{centering}
\end{figure}

In \Fig{fig:ground} we show our numerical lower bounds of the ground state energy, and compare with some well-known points. For the regions with an exact product-state ground state, the MED is able to faithfully represent these and predicts the correct results. In more entangled regions, the results simply lower bound the possible allowed energies --- for example the Monte Carlo results included in \Fig{fig:ground} (accurate to many significant figures) lie above the MED prediction. The difference of a few percent is not surprising, given that a clusters of up to just 8 sites were used with no finite-size scaling analysis.

\begin{figure*}[t]
    \begin{centering}
    \includegraphics[width=17cm]{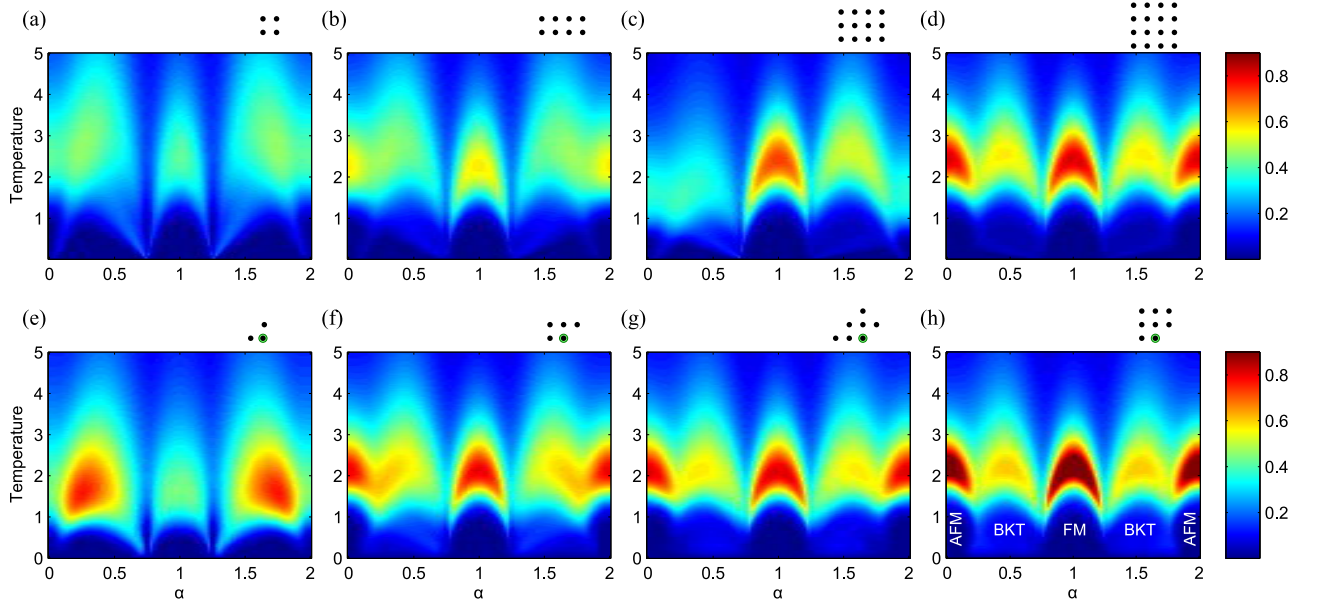}
    \caption{(Color online) Specific heat as a function of temperature and the parameter $\alpha$. (a--d)~Exact diagonalization results with (a)~2$\times$2 lattice, (b)~4$\times$2 lattice, (c)~4$\times$3 lattice, and (d)~4$\times$4 lattice. (e--h)~MED results using a cluster containing (e)~3-sites, (f)~5-sites, (g)~7-sites, and (h)~8-sites. We see from (c) that exact-diagonalization is very sensitive to system size and boundary conditions, while a variety of cluster shapes can be used with the MED. \label{fig:Cv}}
    \end{centering}
\end{figure*}

The specific heat of a system can give an excellent indication of the phase diagram. In \Fig{fig:Cv} we plot the specific heat at different values of $\alpha$ and temperature $T$, calculated by both MED and exact diagonalization with various cluster/system sizes. We observe that the MED produces the general features and sharpness of exact diagonalization simulations using far fewer sites (approximately half). The expected features of the phase diagram are already clearly visible for the exact-diagonalization using 16 (4$\times$4) sites, and MED using just 7 sites.

%\begin{figure*}[t]
%    \begin{centering}
%    \includegraphics[width=4.5cm]{XXZ_E_ED22.eps}
%    \includegraphics[width=4.5cm]{XXZ_E_ED24.eps}
%    \includegraphics[width=6cm]{XXZ_E_ED44.eps}
%    \includegraphics[width=4.5cm]{XXZ_E_MED3s.eps}
%    \includegraphics[width=4.5cm]{XXZ_E_MED5s.eps}
%    \includegraphics[width=6cm]{XXZ_E_MED7s.eps}
%    \caption{(Color online) Energy per particle as a function of temperature and the parameter $\alpha$. (a-c) Exact diagonalization results with (a) 2$\times$2 lattice, (b) 4$\times$2 lattice, and (c) 4$\times$4 lattice. (d-f) MED results using a cluster containing (d) 3-sites, (e) 5-sites, and (f) 7-sites. NOTE: upper plots have problem different zero-energy points. \label{fig:energy}}
%    \end{centering}
%\end{figure*}

In \Fig{fig:Cv} we can clearly observe distinct boundaries between the high-temperature paramagnetic phase, the ferromagnetic (FM) and antiferromagnetic (AFM) discrete broken-symmetry phases, as well as the Berezinskii-Kosterlitz-Thouless (BKT) regions corresponding to systems with quasi-long range order (analogous to the classical rotor model in 2D). At the phase boundary between the high-temperature and (anti)ferromagnetic phases, we expect to observer a narrow, divergent peak in the specific heat; however this is smoothed somewhat by finite-size (or finite-cluster) effects.

%We have marked the maximum of the specific heat in \Fig{fig:Cv} (TODO)

\begin{figure}[t]
    \begin{centering}
    \includegraphics[width=8cm]{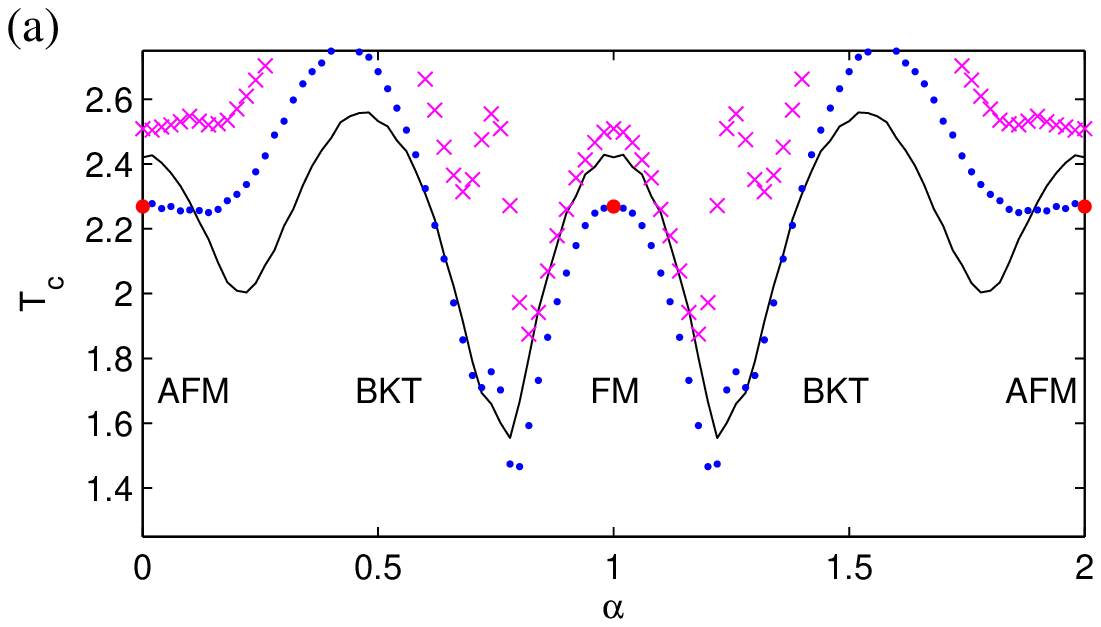}
    \includegraphics[width=8cm]{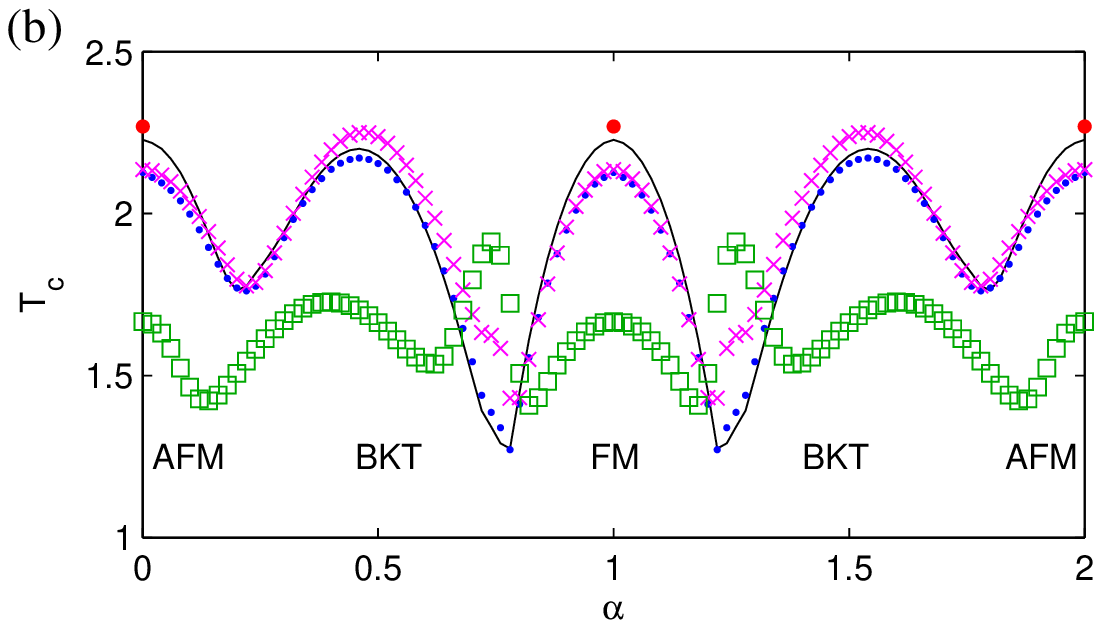}
    \caption{(Color online) Critical temperature as a function of the parameter $\alpha$, extracted from the peak of the specific heat. (a) Results from exact diagonalization, using systems sizes of 4$\times$4 (black, solid line), 4$\times$2 (blue points) and 2$\times$2 (magenta crosses). (b) Results from MED for clusters containing 8-sites (black, solid line), 7-sites (blue points), 5-sites (magenta crosses) and 3-sites (green squares). Red points correspond to the analytic Ising model solution, $T_c = 2/\ln(1+\sqrt{2})$. \label{fig:Tc2}}
    \end{centering}
\end{figure}

The maximum of the specific heat gives a first approximation to the location of the phase transition, and we have plotted this critical temperature in \Fig{fig:Tc2}. The position of the peak is strongly affected by the number of sites in both the very small exact diagonalization systems and small MED clusters. This effect can be most easily quantified at the points $\alpha = 0$ or $1$, where the known value of $T_c = 2/\ln(1+\sqrt{2}) \approx 2.269$~\cite{Onsager1944} is 6.7\% smaller than the 4$\times$4 ED value of 2.421, and 1.8\% larger than the 7-site MED value of 2.228. This analysis fails closer the the SU(2)-invariant quantum critical points ($|\Delta|=1$, or $\alpha = 0.25, 0.75, 1.25, 1.75$) where the narrow peak is obscured by finite-size effects and the broad, smooth maximum expected at higher temperatures. The continuous, BKT transition is generally more challenging to locate, especially when dealing with extremely small systems or, in our case, small MED clusters.

There exist a myriad of more sophisticated, finite-size scaling techniques for determining the critical temperature~(e.g. those employed in~\cite{Sandvik1997,Sandvik1999}). One popular finite-size technique is to investigate the crossing of the specific heat at different system sizes; in general these crossings will approximately coincide and converge rapidly to critical point with increasing system size $L$. Unfortunately, both exact-diagonalization and MED give poor results for these system sizes --- in \Fig{fig:Cv} we observe the smaller simulations don't even clearly recover four distinct low-temperature phases, and we do not have enough data to perform the crossover analysis. If one were to apply this approach using larger clusters, it is apparent that care must be taken with cluster boundaries (i.e. to compare clusters with similar shaped neighbourhoods $\mathcal{N}_k$), just as similar boundary conditions should be used in scaling analysis of exact-diagionalization or Monte Carlo results. One further complication results from the MED method itself --- the fact that $F$ increases monotonically with larger cluster sizes adds constraints to its derivatives (i.e. the energy and specific heat) that potentially complicate the crossover analysis. A fuller understanding of scaling in the MED is the subject of future work.

%\begin{figure}[t]
%    \begin{centering}
%    \includegraphics[width=9cm]{XXZ_Tc_ED3.eps}
%    \includegraphics[width=9cm]{XXZ_Tc_MED3.eps}
%    \caption{(Color online) Critical temperature as a function of the parameter $\alpha$, extracted using the crossover method. Upper: results from exact diagonalization, comparing the 4$\times$4 and 4$\times$2 cells. Lower: results from MED, comparing the 7-site and 3-site clusters. Could combine curves onto single plot??\label{fig:Tc}}
%    \end{centering}
%\end{figure}

\begin{figure}[t]
    \begin{centering}
    \includegraphics[width=\columnwidth]{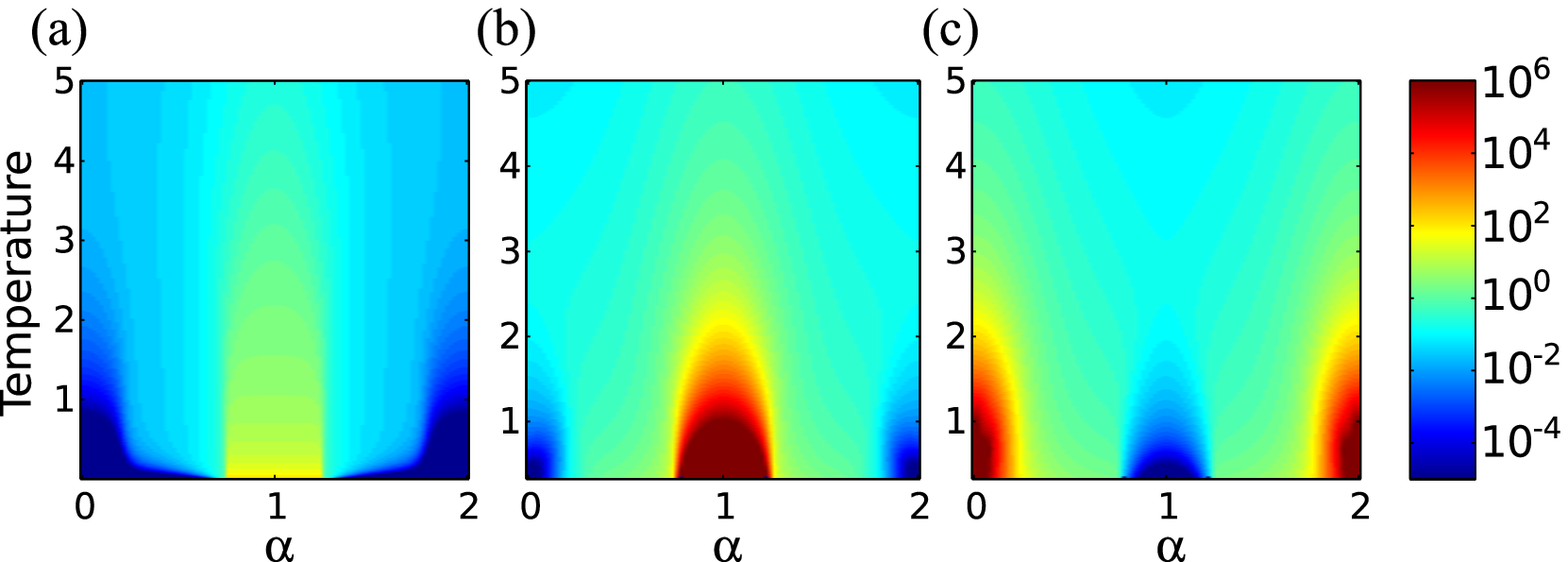}
    \caption{(Color online) Magnetic susceptibility as a function of temperature and the anisotropy parameter $\alpha$. (a)~Susceptibility in the 4$\times$4 lattice. (b)~Susceptibility given by MED with an 8-site cluster. (c)~Susceptibility of the staggered magnetization given by MED with an 8-site cluster. Note the logarithmic color-scale.
    \label{fig:chi}}
    \end{centering}
\end{figure}

Finally, we can extract information on the behaviour of each distinct phase by investigating susceptibility. Although neither MED nor exact-diagonalization of finite-systems displays explicit symmetry breaking, the onset of long-range order in the thermodynamic limit is predicated by rapidly growing susceptibility with system size, caused by strongly-cooperative many-body behaviour. In \Fig{fig:chi} we show the magnetic susceptibility calculated by exact-diagonalization and the MED. In both cases, we can see that the magnetic susceptibility grows extremely rapidly in the ferromagnetic region, while being suppressed in the antiferromagnetic phase. However, the many-body behaviour is much stronger in the MED, where the susceptibility is orders of magnitude greater. The same analysis can be applied, for instance, to the staggered magnetization to characterize the anti-ferromagnetic region (\Fig{fig:chi}~(c)), giving results as expected.

\section{Conclusion}

\label{conclusion}

The Markov entropy decomposition is flexible cluster-based method for finite temperature quantum calculations in any dimension or geometry, giving rigorous lower bounds to free- and ground-state energies. In this paper we have detailed efficient algorithms for optimizing the MED clusters and demonstrated its applicability to the XXZ model on the square lattice. We observe that MED simulations share many properties with exact diagonalization, both having exponential scaling in cost with cluster size, strong finite-size effects, and the lack of explicit phase transitions or symmetry breaking. On the other hand, the MED approach is much more accurate than exact-diagonalization for a given number of lattice sites, displays stronger many-body effects, and further is much more flexible in the types of geometries that are feasible to study. For example, one could study large, finite systems (e.g. harmonically-trapped atoms in optical lattices), systems with large unit cells (e.g. the hyperkagome lattice), or investigate symmetry breaking over large unit cells or that are incommensurate with the lattice (e.g. systems with chiral order). Further, the close relationship with belief propagation~\cite{Hastings2007} leads us to believe that this method would be excellent for the study of disordered, glassy systems at finite-temperature, where correlation lengths are typically short, but the relevant global states are difficult to determine. The authors hope that this paper stimulates further work with this promising method.

\subsection*{Acknowledgements}

This work was supported by  NSERC and FQRNT through the network INTRIQ. Computational resources were provided by Compute Canada and Calcul Quebec.

\section*{Appendix}

In this Appendix, we derive the form the Hessian (Eqs.~(\ref{hessian0}--\ref{hessian2})) using matrix calculus. We begin with the unconstrained problem, ignoring the projector $\mathcal{P}$. The first-derivative in the unconstrained subspace is given by \Eqn{dF}, restated here.
\begin{equation}
    \frac{d\tilde{F}}{d\hat{\rho}_i} = \hat{H} + k_B T \left ( \log \hat{\rho}_i - \frac{\hat{\mathbb{1}}}{d} \otimes \log \hat{\rho}^{\prime}_i \right) + \mu \hat{\rho}_i^{-1} \label{dF_appendix}
\end{equation}

Before continuing, lets examine matrix functions. A scalar function $f(x)$ can be converted to the matrix function $f(\hat{x})$ by performing it's action on the eigenvalues of $\hat{x}$. That is, if (Hermitian) $\hat{x}$ has eigenvalue decomposition $\hat{U}\hat{D}\hat{U}^{\dagger}$, then $f(\hat{x}) = \hat{U}\hat{E}\hat{U}^{\dagger}$ where $f(\cdot)$ is applied element-wise to $\hat{D}$:
\begin{equation}
  \hat{E} = \left[ \begin{array}{ccc} f(D_{1,1}) & 0 & \cdots \\ 0 & f(D_{2,2}) & \cdots \\ \vdots & \vdots & \dots \end{array} \right].
\end{equation}
This is in-fact the meaning behind the terms in \Eqn{dF_appendix}.

Next we wish to examine how the matrix $f(\hat{x})$ changes if an element of $\hat{x}$ changes. To simplify this matrix derivative, we will proceed to work in the basis that $\hat{x}$ is diagonal. First we examine what happens when a small change occurs to a diagonal element.
\begin{eqnarray}
    \frac{d E_{kl}}{d D_{ii}} & = & \lim_{h\rightarrow 0} \delta_{kl} \left( f(D_{kk} + h\delta_{ik}) - f(D_{kk}) \right) / h \nonumber \\
    & = & \delta_{kl} \delta_{ik} f^{\prime}(D_{ii})
\end{eqnarray}
Note $f^{\prime}$ is the derivative of $f$. Unsurprisingly, we get $dE_{ii}/dD_{ii} = f^{\prime}(D_{ii})$.

The off-diagonal terms are slightly more complicated. Adding a small perturbation to an off-diagonal element $D_{ij}$ leaves the matrix diagonal everywhere except the $2\times2$ sub-matrix involving rows and columns $i$ and $j$. The problem can be solved by dealing just with this sub-matrix,
\begin{eqnarray}
    \frac{d E_{ \{i,j\},\{i,j\} }}{d D_{ij}} & = & \lim_{h\rightarrow 0}  \frac{ f\left(\left[ \begin{array}{cc} D_{ii} & h \\ 0 & D_{jj} \end{array} \right] \right) -  f\left(\left[ \begin{array}{cc} D_{ii} & 0 \\ 0 & D_{jj} \end{array} \right] \right) }{h} \nonumber \\
    & = &  \left[ \begin{array}{cc} 0 & \frac{f(D_{ii}) - f(D_{jj})}{D_{ii} - D_{jj}} \\ 0 & 0\end{array} \right] \label{offdiag}
\end{eqnarray}
The derivative is zero otherwise. In the limit $D_{ii} \rightarrow D_{jj}$, the term approaches $f^{\prime}(D_{ii})$. A simple way to see this result is to prove it for all functions $f(\hat{x}) = \hat{x}^n$,
\begin{eqnarray}
    \frac{d E_{ \{i,j\},\{i,j\} }}{d D_{ij}} & = & \lim_{h\rightarrow 0}  \frac{ \left[ \begin{array}{cc} D_{ii} & h \\ 0 & D_{jj} \end{array} \right]^n -  \left[ \begin{array}{cc} D_{ii} & 0 \\ 0 & D_{jj} \end{array} \right]^n }{h} \nonumber \\
    & = &  \left[ \begin{array}{cc} 0 & \frac{D_{ii}^n - D_{jj}^n}{D_{ii} - D_{jj}} \\ 0 & 0\end{array} \right].
\end{eqnarray}
Combining this with a simple Taylor expansion of $f(\hat{x})$ generalizes this to \Eqn{offdiag}. (The result holds for all differentiable $f$, regardless of the radius of convergence of the Taylor expansion).

The tensor $d\hat{E}/{d\hat{D}}$ is actually a diagonal, linear super-operator whose action is element-wise multiplication by the matrix $\hat{M}$, where
\begin{equation}
   \hat{M} = \left[  \begin{array}{ccc} f^\prime(D_{1,1}) & \frac{f(D_{1,1}) - f(D_{2,2})}{D_{1,1} - D_{2,2}} & \cdots \\ \frac{f(D_{1,1}) - f(D_{2,2})}{D_{1,1} - D_{2,2}} & f^{\prime}(D_{2,2}) & \cdots \\ \vdots & \vdots & \ddots \end{array}  \right]. \label{matrix_derivative}
\end{equation}

To apply this operator, we must first transform to the basis where $\hat{x}$ is diagonal, by multiplying on the left by $\hat{U}$ and on the right by $\hat{U}^{\dagger}$. The super-operator that performs this operation is $\hat{U} \otimes \hat{U}^{\ast}$. We then multiply element-wise by $\hat{M}$, or rather apply the diagonal super-operator $\text{diag} \, \hat{M}$, before performing the inverse unitary transform $\hat{U}^{\dagger} \otimes \hat{U}^{T}$. Together, this becomes
\begin{equation}
    \frac{d f(\hat{x})}{d \hat{x}} = \hat{U}^{\dag} \otimes \hat{U}^{T} \cdot \mathrm{diag} \; \hat{M} \cdot \hat{U} \otimes \hat{U}^{\ast}.
\end{equation}
This now explains the first line of \Eqn{hessian1}. Equations~(\ref{hessian1a},\ref{hessian1b}) come from inserting $f(\hat{x}) = k_B T \log \hat{x} + \mu \hat{x}^{-1}$ into \Eqn{matrix_derivative}.

Finally, a simple extension is required to deal with the reduced density matrices $\hat{\rho}^{\prime}_k$. First we split the Hilbert space into two components, that relating to site $k$, and the remainder of the cluster $\mathcal{M}^{\prime}_k$. We define the (linear) super-operator that performs the partial trace over site $k$ as $\frak{pt}$, so that $\frak{pt} \, \hat{\rho}_k = \hat{\rho}^{\prime}_k$. We define the quasi-inverse of this operator, the `expansion' super-operator, that adds the fully mixed state on site $k$, as $\frak{ex}$, where $\frak{ex} \, \hat{\rho}^{\prime}_k = \hat{\rho}^{\prime}_k \otimes \hat{\mathbb{1}}/d$. The $\hat{\rho}^{\prime}_k$ term in \Eqn{dF_appendix} becomes
\begin{equation}
   -k_B T \, \frak{ex} \left\{ \log\left( \frak{pt} \left\{ \hat{\rho}_k \right\} \right) \right\}.
\end{equation}
To differentiate this composition, we simply apply the standard chain rule,
\begin{equation}
   \frac{d \frak{ex} \{ \hat{z} \}}{d\hat{z}} \; \frac{d \log ( \hat{y} )}{d\hat{y}} \; \frac{d \frak{pt} \{ \hat{x} \}}{d\hat{x}},
\end{equation}
where $\hat{y} = \frak{pt}\, \hat{x}$ and $\hat{z} = \log \hat{y}$. The linear super-operators $\frak{ex}$ and $\frak{pt}$ are equal to the derivative of their action (this is true for any linear super-operator), so this simplifies to
\begin{equation}
   \frak{ex} \cdot \frac{d \log ( \hat{y} )}{d\hat{y}} \cdot \frak{pt} .\label{chain_rule2}
\end{equation}
The central derivative can be defined in the basis where $\hat{\rho}^{\prime}_k$ is diagonal, in much the same way as above. This leads directly to the second term in \Eqn{hessian1}.

\bibliography{../bib/andy.bib}

%\begin{thebibliography}{99}

%DMRG
%\bibitem{White1992}
%S.R. White, Phys. Rev. Lett. {\bf 69}, 2863 (1992).

%\end{thebibliography}

\end{document}